# Formation of tiny particles and their extended shapes –Origin of physics and chemistry of materials

**Mubarak Ali [a, *] and I-Nan Lin [b]**

[1]Department of Physics, COMSATS University Islamabad, Park Road, Islamabad-45550, Pakistan; *E-mail: mubarak74@mail.com or mubarak74@comsats.edu.pk

[b] Department of Physics, Tamkang University, Tamsui, New Taipei City 25137, Taiwan

**Abstract** – Tiny-sized particles under the scheme of monolayer assembly, comprising gold atoms, developed at a different processing time in a pulse-based process. For a different processing time, atoms bind into different tiny particles under the placing packets of nanoshape energy where they elongate as per arrangement and when in one-dimensional arrays, they convert into structures of smooth elements. For different processing time and where tiny particles possess triangular-shape, they pack to develop extended shapes where development rate of an anisotropic particle is not more than millisecond time. Increasing the processing time of solution upto certain duration increases the number of developing tiny particles in a triangular-shape, so, their extended shapes also. Uniformly adjacent-orientation of electrons in atoms of tiny-shaped particle is because of exerting uniform surface force along their opposite poles as per gained potential energy where stretching of their clamped energy knots is remained orientational-based. At a different processing time, inter-spacing distance of spotted intensity spots in selective area photons reflection patterns of particles is remained the same as for the case of their structures of smooth elements visualized through transmission microscope high-resolution images. When the forceful coinciding of two parallel structures of smooth elements is occurred, they bind into single element structure (of smooth element) by a bit overlying inner sides, thus, giving its double width where certain filled state electrons and unfilled energy knots (belonging to sides of elongated atoms of parallel structures of smooth elements) coordinate to adhere. This study discusses the formation of tiny particles following by their extended shapes at different processing time of gold solution while employing a pulse-based electron-photon solution-interface process where they become the origin of physics and chemistry of materials by discussing many commonly-known phenomena and processes, so, opening the alternative routes to design materials and explore science.

*Keywords:* Process time; Tiny particles; Particles; Electron-dynamics; Diffusion; Electricity and magnetism; Force-energy

1. Introduction

To synthesize particles comprising certain nature of atoms at a different processing time is easy but it is difficult to understand underpinning mechanisms of their development. In tiny-sized particles, inherent nature of atoms should be considered (taken) as the source of forming their shape and structure and then the details of their processing conditions. To develop highly-shaped colloidal particles having high aspect ratio is the way forward to advance science, hence, their sustainable technology. Development of new synthesis processes for those diversified-class materials may open new avenues where the observation of their hidden features (at sub-atomic) could be possible, thus, their studies provide the opportunities to think beyond the existing knowledge. At first stage, this is possible by studying different-class metallic colloids. However, to explore the mechanisms of developing different tiny-metallic colloids and their assembling into large-sized particles, it is vital to study them in a order where our thoughts might build a solid-connection to originate their science.

Several studies in the literature are enlisted explaining different approaches to synthesize metallic colloids along with their development mechanisms and applications where duration of the process is appeared to be an important parameter. But, it is being seldomly considered for their processing. Anisotropic gold particles were synthesized at the solution surface while distorted ones were synthesized inside the solution [1]. Tiny-metallic colloids are the suitable candidates in the new emerging applications of catalysis [2, 3]. Catalytic activity of metal nanostructures is enhanced by controlling the phase [4, 5]. On-going research efforts not only consider geometry and entropy to explain the structure but the dynamics also [6]. Some metrics possess order which precisely characterize their packing [7].

Since six decades, a lot of work has been carried out on the localized surface plasmons where it was the result of the confinement of a surface plasmon in a nanoparticle of size comparable to (or smaller than) the wavelength of light used to excite the plasmon. However, surface plasmon polaritons travel along a metal-dielectric or metal-air interface and the wave involves both surface plasmons (charge motion in the metal) and polaritons (electromagnetic waves in the air or dielectric) [8]. Plasmons offer an unusual capability to confine light to the subwavelength scale [9]. Again, in physical chemistry, van der Waals interactions are the residual attractive or repulsive forces between molecules or atomic groups that don't arise from a covalent bond or electrostatic interaction of ions or of ionic groups with one another or with neutral molecules [10]; the term includes dipole-dipole, dipole-induced dipole and London (instantaneous induced dipole-induced dipole) forces. The resulting van der Waals forces can be attractive or repulsive [11]. Between two isolated Rydberg atoms, direct measurement of the van der Waals interaction is reported which opens exciting point of views in multiatomic systems [12]. Kawai *et al*. [13] studied van der Waals interactions in rare gas atoms and highlighted the roles in an isolated atomic model while identifying the limits. The role of van der Waals interactions in the binding of small molecules in the gas phase is clear [14]. However, there are numerous reports available in the literature discussing and explaining the light-matter interactions. Again, diffusions of matter at the scale of atoms and grains have also a long history and their mechanisms have been continuously explained by the established fick's laws along with other available theories. This also raises a question on the status of science behind magnetic materials.

In this study, it is discussed that the formation of tiny-sized particles along with their extended shapes under varying the processing time of gold solution is related to different mechanism of their underlying science than the available ones. By taking into consideration their underlying mechanism of development, specifically, in the case of formation of tiny-sized particles at a different time of processing their solutions, it is inferred that binding of atoms for tiny colloids don't fulfil the descriptions falling under van der Waals interactions. The formation mechanism of a tiny particle partially supports the theory where only the element of force is not enough. Binding of atoms also requires the energy. This study goes into deeper-side explaining that trapping of light by metallic tiny-cluster while travelling along the matter-solution interface doesn't result into the collective oscillations of its atoms. But, atoms of tiny-sized particle at solution surface are affected by the forcing energy of travelling photons (light) in different ways; in addition to interaction of photons to atoms, the existing state of the atom is also considered. At surface of solution, atoms of tiny-sized particles remain in their certain transition state because of decreasing the force of their grounded format (original solid behavior); electrons of their certain transition state atoms are in the grip of surface forces also instead of more for the grip of grounded format (south-pole) force. The relation of force (influencing/exerting) at electron-level with atom (and tiny particle) at different level of ground originates the science of many new phenomena, where some of them are known in diffusion processes, electricity and magnetism principles and applications along with related to general physics and chemistry. Thus, the formation of tiny particles and their extended shapes set the origin of physics and chemistry of materials

In this work, a combined approach is employed to understand the development process of gold tiny-sized particles following by large-sized particles (extended shapes) at a different time of processing their solutions when the fixed precursor concentration was chosen for each experiment. Origin of attaining adjacent-orientation of electrons belonging to outer rings of gold atoms forming tiny-metallic colloids as per exerting surface forces is discussed. Under the forceful coinciding of two parallel structures of smooth elements, their convertion into one under the maintenance of double width is discussed. Diffusion of electrons (of atoms of tiny particles) by having certain orientation under the orientational-based stretching of their clamped energy knots is described where they discuss the origin of physics and chemistry of materials in many ways and in different aspects to the available ones.

## 2. Experimental details

HAuCl$_4$.3H$_2$O, ACS, 99.99 % (metals basis), Au 49.5 % min crystalline was purchased from Alfa Aesar. A fixed precursor concentration 0.60 mM was chosen for each experiment to process solution for 2 minutes, 15 minutes and 20 minutes. Total quantity of solution for each experiment was 100 ml. A 10 µsec pulse ON/OFF time under bipolar polarity was chosen in each experiment. Further detail of preparing gold solution along with schematic of air-solution interface and electron-photon-solution interface is given elsewhere [15]. However, a complete setup of the process is discussed in another study [16]. For bright field and high-resolution images of transmission microscopy, samples were prepared by pouring a drop of processed solution, at a different time, on carbon-coated copper grid. Bright field images of gold particles were taken by transmission microscope known as TEM (JEOL JEM2100F, operated at 200 kV) while high-magnification view was captured by high-resolution transmission microscope. Selected area photon reflection (SAPR) patterns of prominent particles developed at a different processing time of their solution were investigated under the application of transmission microscopy. The SAPR pattern is referred to SAED (selected area electron diffraction) pattern in the literature but the recently published studies agreed that it is a photon reflection and not electron diffraction [22, 24]. Surface features (shape or morphology) of nanoparticles and particles were also examined by field emission scanning microscopy (FE-SEM; ZEISS-SIGMA) in the case of solutions processed at 15 minutes and 20 minutes process time where polished surface of silicon wafer was used to adhere the drops of prepared solutions. After drying the dropped solution at surface of silicon wafer, samples were used to examine the shape (morphology) of particles having different extended shapes.

## 3. Results and discussion

For the shortest processing time of solution, the gold particles show mixed-trend of developing as their bright field transmission microscopy images (a-h) are shown in Figure 1; in some cases, the tiny-shaped particles packed at the precise unfilled regions resulting into the faceted and smooth shapes of particles (in Figure 1a-c and pentagonal-shaped particle in Figure 1e); in some cases, the tiny-sized particles partially packed to develop particles of extended shapes as shown in Figure 1 (d) and in Figure 1 (e) other than the pentagonal-shaped particle. However, in Figure 1 (f) and Figure 1 (g), the tiny-sized particles are recognized in the developed particles of unfaceted extended shapes of triangle and hexagon. In Figure 1 (h), the particle indicates unexploited packing of tiny-sized particles along the two sides of developing shape, most likely to be in a triangular-shape. In the distorted particles, the packing of tiny-sized particles remained out of order mainly due to geometric constraint. In some particles, tiny-sized particles don't pack to fill the regions of developing particles in a fit manner where their boundaries are clear to observe as in the case of Figure 1 (f) and (g).

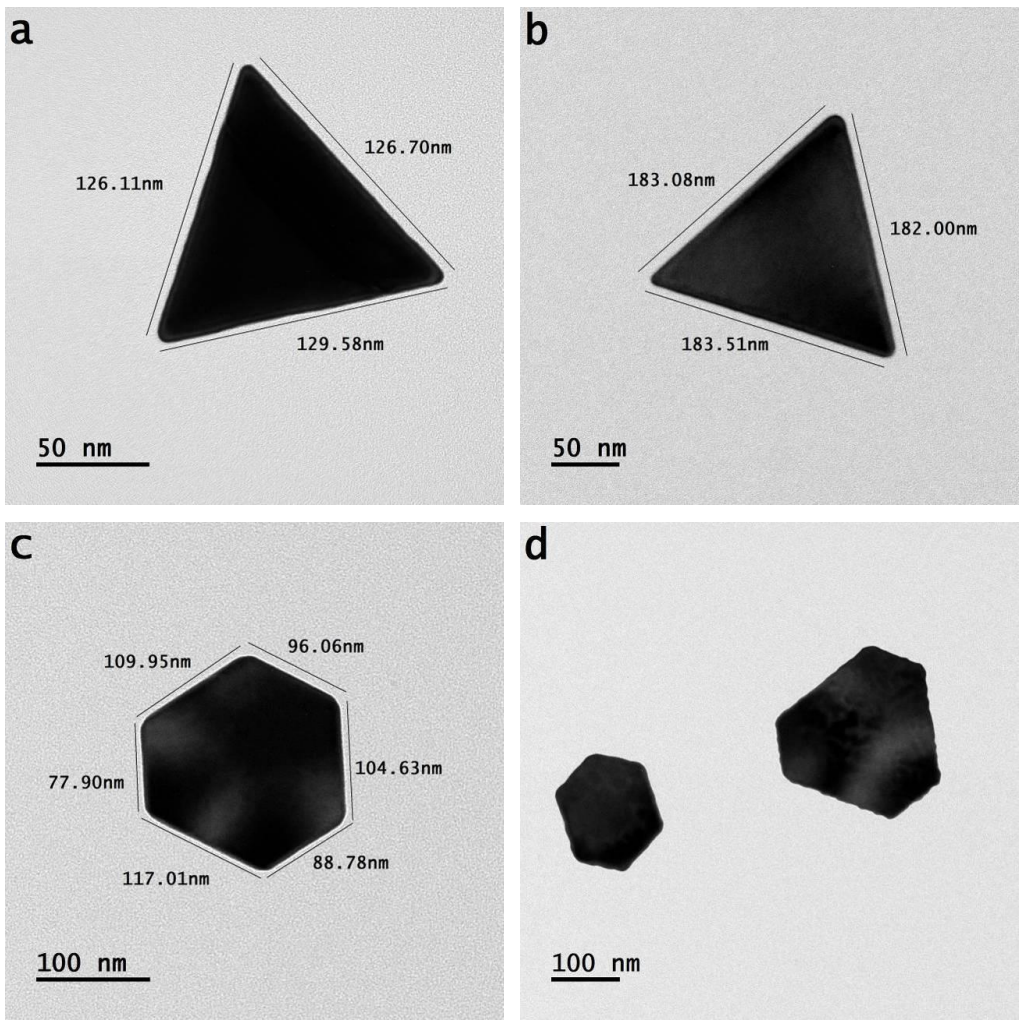

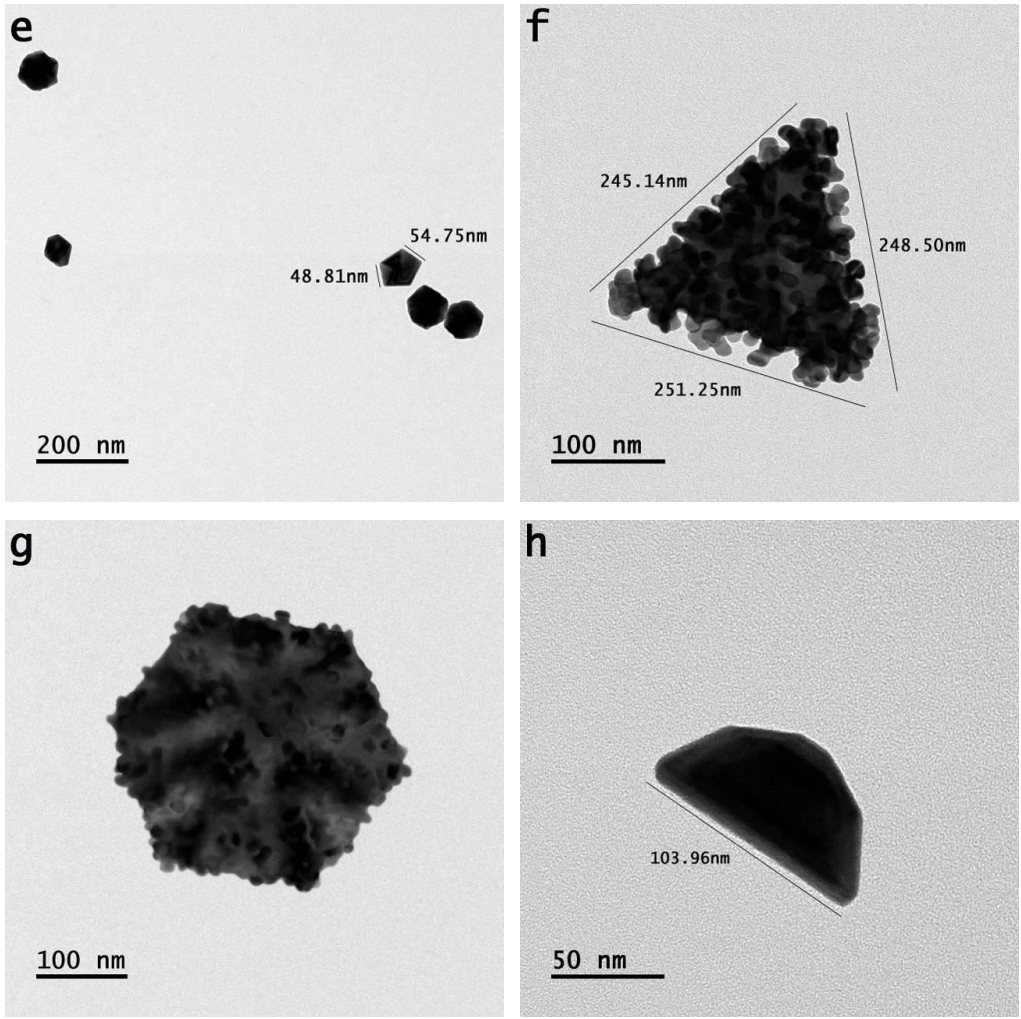

**Figure 1:** Bright field transmission microscope images (a-h) of gold particles synthesized at a processing time of 2 minutes

For 15 minutes processing time of solution, the resulted extended shapes of particles don't show any sign of unfit packing of tiny-sized particles and they packed precisely to fill unfilled regions, that is why, their coalescence is termed as packing, because of giving the impression and gesture like that, which is prior to the assembling of their structures of smooth elements. The extended shapes of the particles are faceted and smooth as shown in bright field microscopy images of Figure 2 (a)-(h). Some of the particles were developed in distorted shapes as shown in Figure 2 (g) and Figure 2 (h) where tiny-sized particles other than triangular-shape packed under the exertion of mixed-behavior forces. In Figure 2 (A) and Figure 2 (B), the SAPR patterns of regular and non-regular hexagonal-shaped particles measure inter-spacing distance of their structures of smooth elements ~0.24 nm; distance between printed intensity dots (center-to-center) of reflected photons in the pattern is uniform throughout the selected region of structure. On joining intensity points in the reflection patterns, a perfect shape of triangle is constructed and distance between any two nearest intensity spots (center-to-center) gives nearly equal length as labeled in the patterns of Figures 2 (A) and (B) indicating the equal distance of reflected photons from the surface of structures of smooth elements. A pentagonal-shaped particle in Figure 2 (d) is smaller in size to the one shown in Figure 2 (f) indicating the packing of smaller size tiny-sized particles to develop their extended shape. Due to the imprecise (inexact) filling of unfilled region of developing particle under the packing of tiny-sized particles, their particle of rhombus shape indicates wrinkles as shown in Figure 2 (c). The same is the case in pentagonal-shaped particles shown in Figure 2 (d) where packing of tiny-shaped particles in the region of five axes resulted into introduce wrinkles (twin boundaries) at the center and packing of tiny-shaped particles couldn't locate a common centre. As shown in Figures 2 (g) and (h), both distorted particles and geometric anisotropic particles developed when 15 minutes of processing time of solution was set.

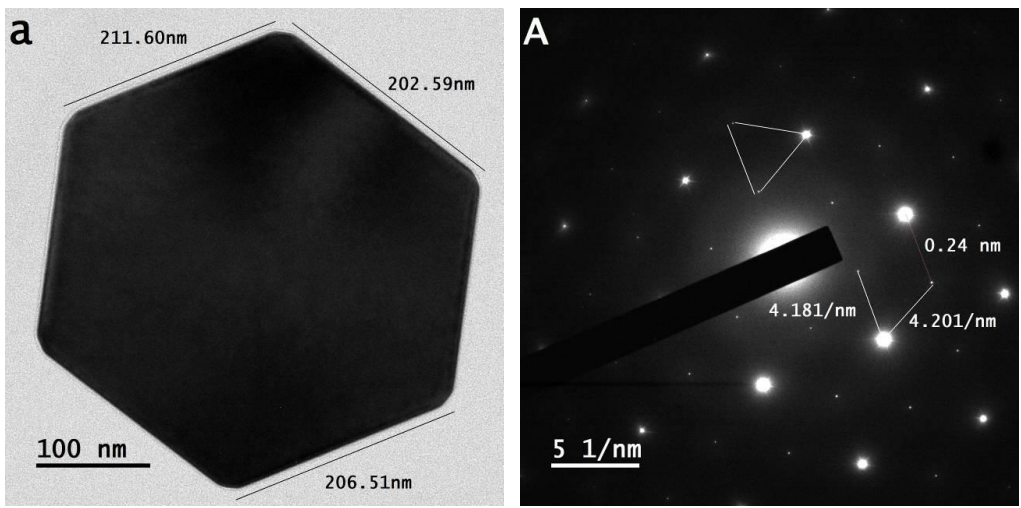

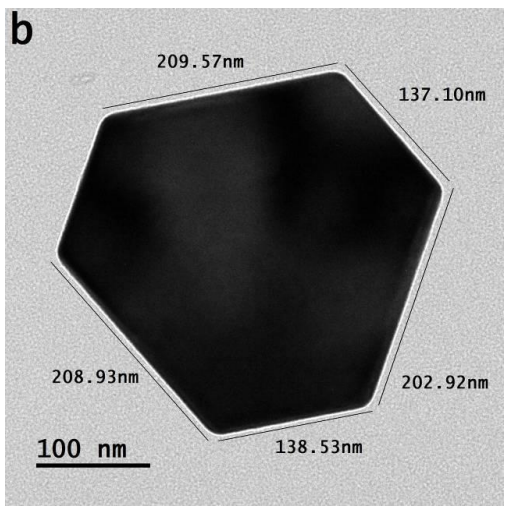
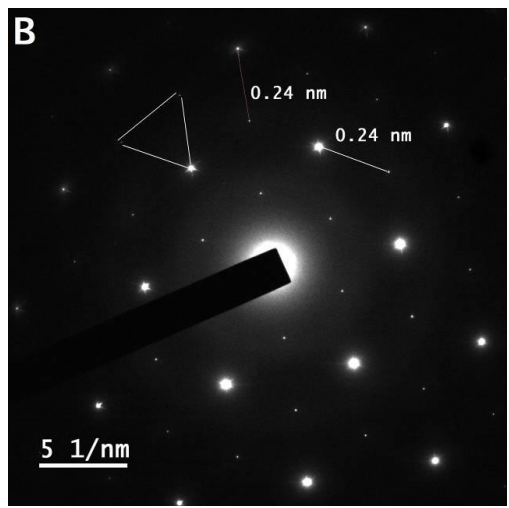
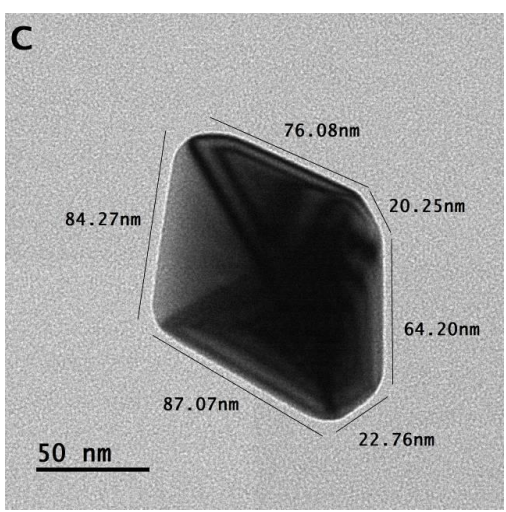
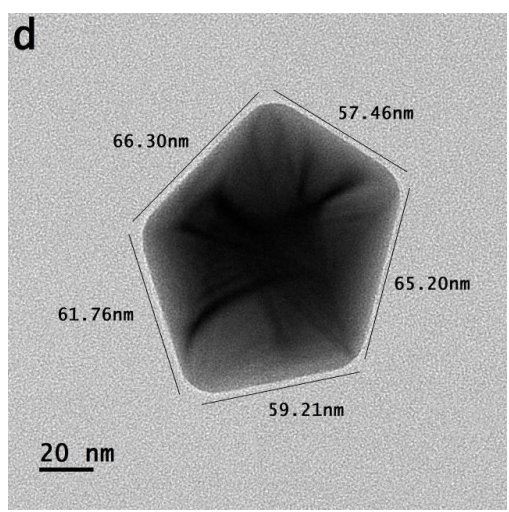
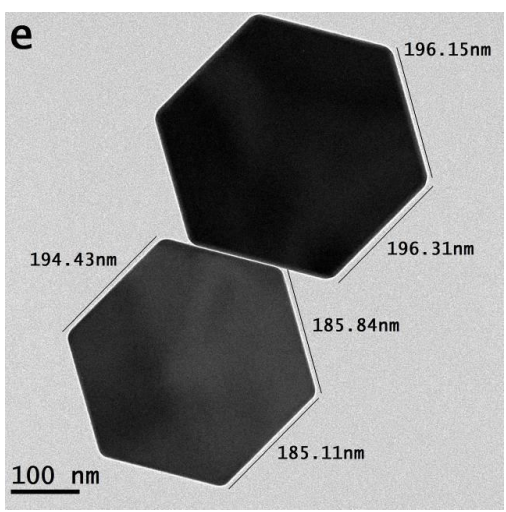
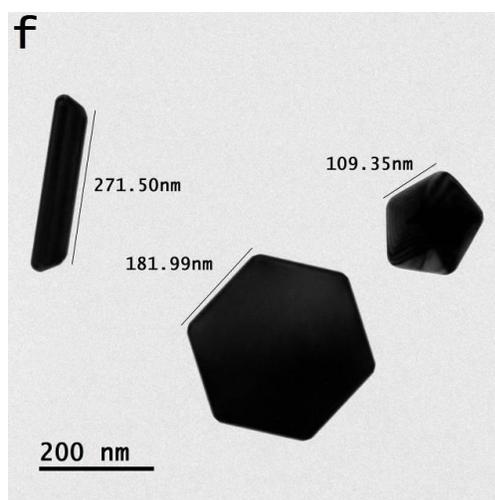

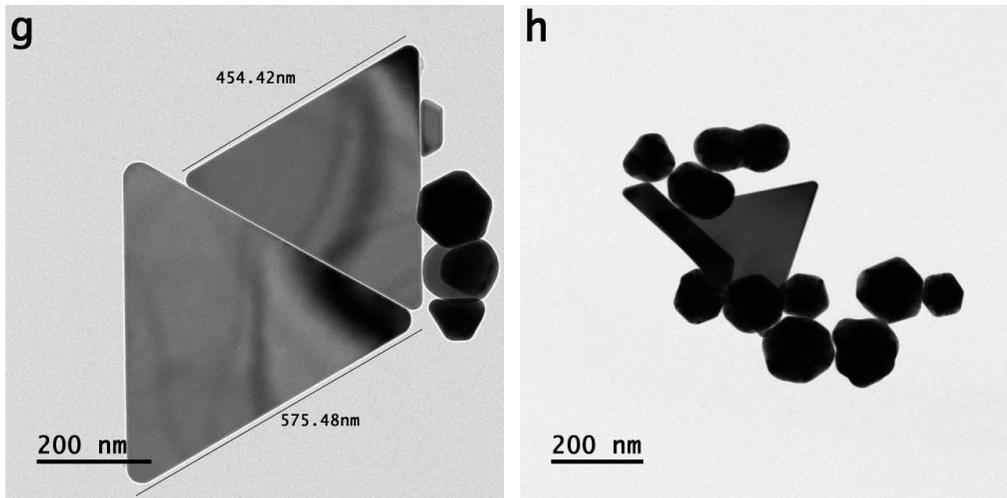

**Figure 2:** Bright field transmission microscope images (a-h) of gold particles synthesized at a processing time of 15 minutes and (A & B) SAPR patterns of hexagonal-shaped particles

In Figure 3 (a-g), different bright field transmission microscope images of gold particles are shown which were developed when the processing time of solution was set 20 minutes. No significant difference in size and shape of particles was observed as compared to the particles synthesized at 15 minutes processing time. In Figures 3 (A-D), SAPR patterns of different hexagonal- and triangular-shaped particles show the same inter-spacing distance of spotted intensity dots (center-to-center) as noted in the case of particles synthesized at a processing time of 15 minutes. However, in Figure 3 (e), the SAPR pattern of rod-shaped particle gives greater inter-spacing distance of parallel spotted lines of intensity spots (mid-to-mid). Selected area photons reflection patterns validate that center-to-center distance of intensity spots (because of forming the dots) is ~0.24 nm for the case of particles having multi-dimensional shapes and mid-to-mid distance of intensity spots (because of forming the lines) is ~0.27 nm for the case of particles having one-dimensional shape.

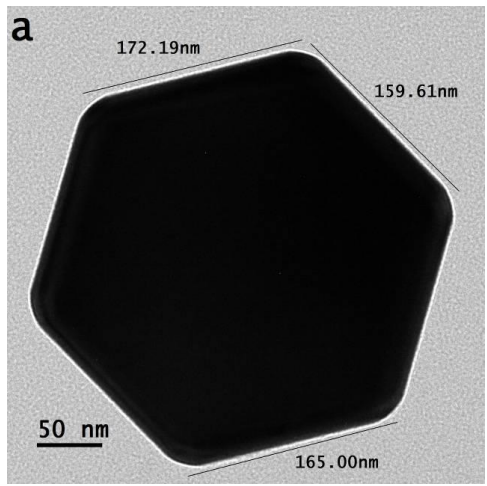
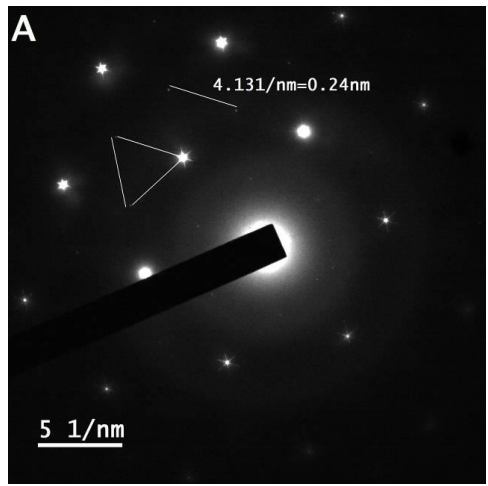
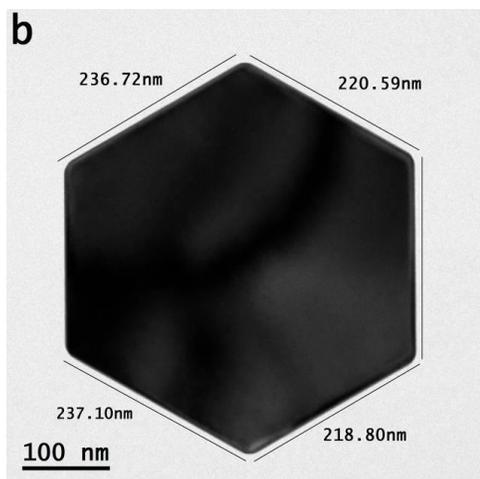
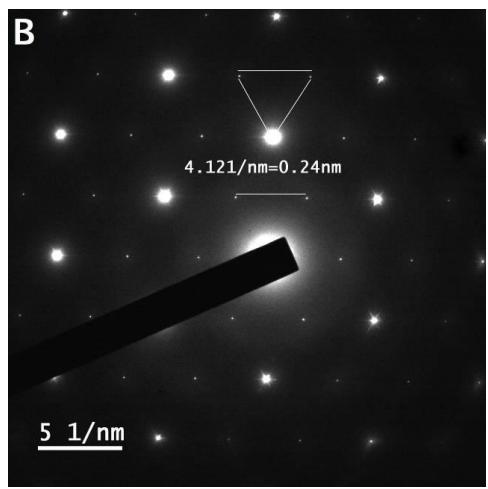
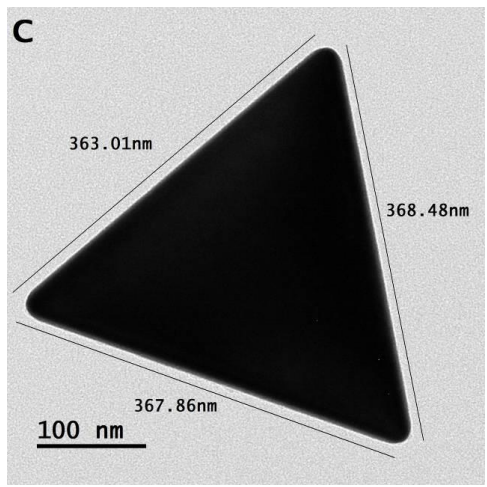
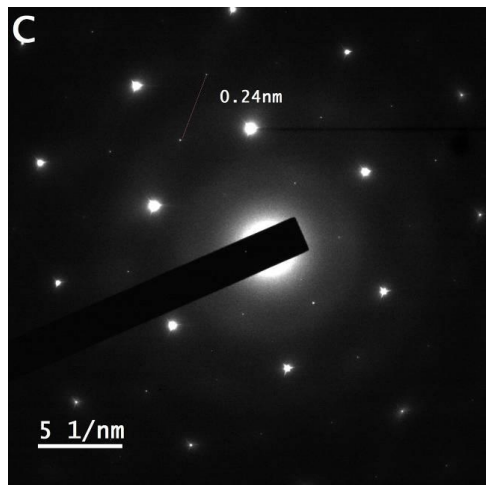

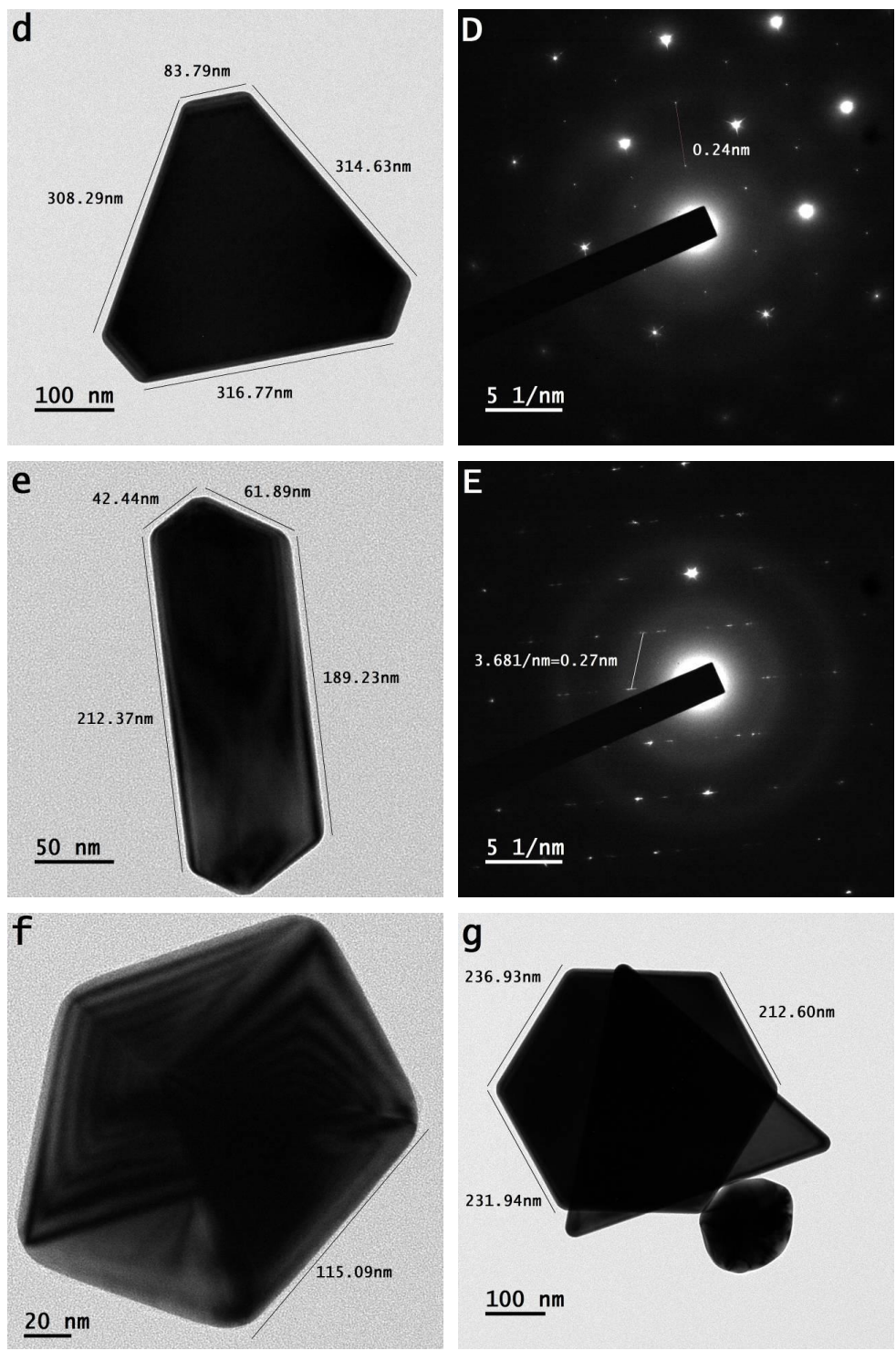

**Figure 3:** Bright field transmission microscope images (a-g) of gold particles synthesized at a processing time of 20 minutes and (A-E) SAPR patterns of hexagonal-, triangular- and rod-shaped particles

In Figure 3 (f), the structure of pentagonal-shaped particle shows stresses under the compensation of filling the regions along the five axes. Again, in Figure 3 (g), some of the shapes of particles overlapped and attached to the distorted ones depending on the modes of packings of their tiny-sized particles.

A pentagonal-shaped particle is shown in Figure 4 (a) where lengths of five sides are nearly equal and more stresses are appeared in the center-portion of the shape. High-magnification view of the region marked with square box is shown in Figure 4 (b) where five distinct regions are labeled; (1) a region where structures of smooth elements along with their inter-spacing distance have the same width (~0.12 nm), (2) a region where atoms of tiny-sized particles elongated less and their electrons undertook stretching of clamped energy knots along different sides, (3) a region where atoms of tiny-shaped particles elongated more and converted structures into structures of smooth elements where each have a twice width (~0.21 nm) as compared to the one shown in region 1 of Figure 4 (b). In Figure 4 (b), a region labelled by (4) shows structure of wrinkles which dealt (undertook) different level of stresses and, in the region (5), atoms didn't modify (convert) their structures to structures of smooth elements. Very high-magnification view of region (3) in Figure 4 (b) has been shown in Figure 4 (c) which highlights the width of each structure of smooth element (and inter-spacing distance of structures of smooth elements also), which is twice to the ones shown in Figure 4 (d) i.e., the one cropped from the region (1) of Figure 4 (b).

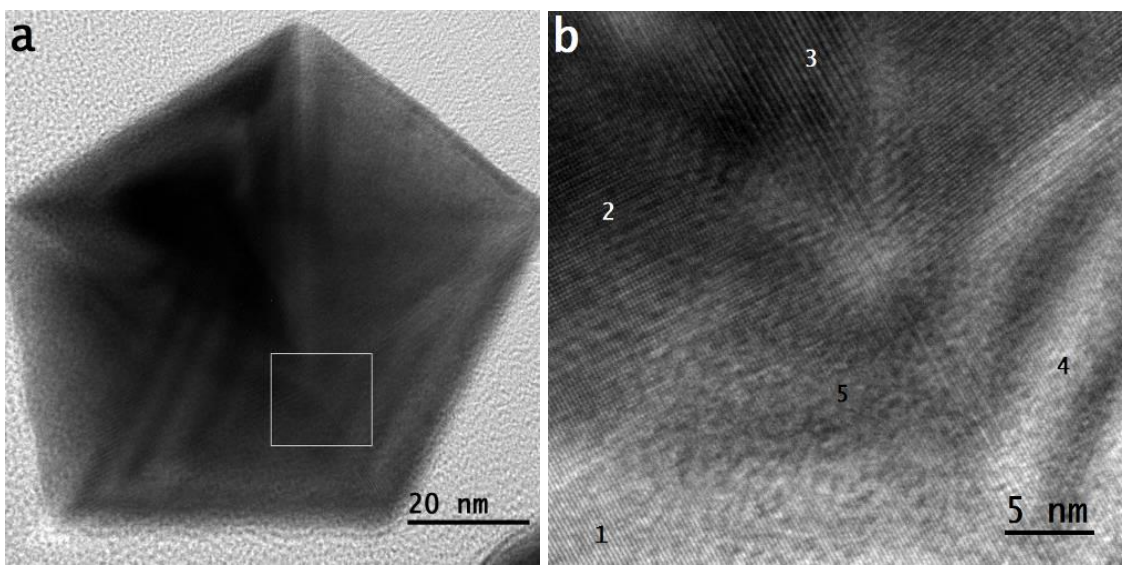

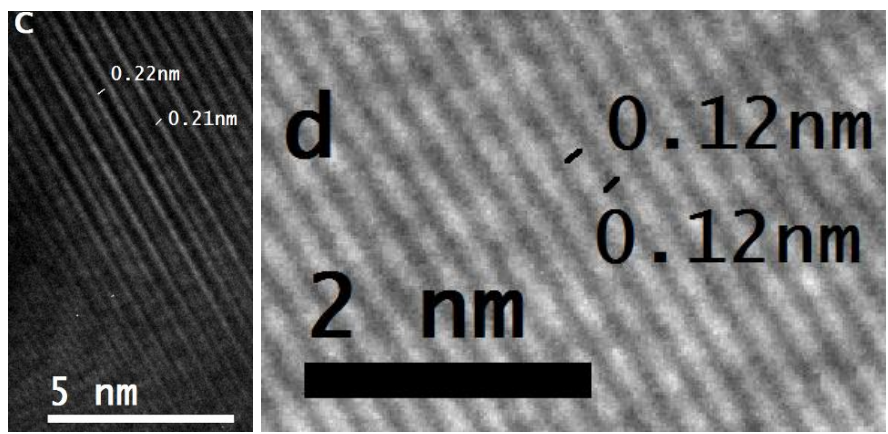

**Figure 4:** (a) bright field transmission microscope image of pentagonal-shaped particle synthesized at a processing time of 2 minutes, (b) high-magnification view of the region enclosed by square box in Figure 4a, further magnified views of the regions labeled by (c) 3 in Figure 4b and (d) 1 in Figure 4b

In Figure 5 and Figure 6, field emission scanning microscope images of gold particles are shown which were synthesized at the processing time of solution 15 minutes and 20 minutes, respectively, indicating the same features of particles having extended anisotropic shapes and distorted shapes as examined in the case of images of particles captured by bright field transmission microscope (shown in Figure 2 and Figure 3) where mainly the difference is in their texture.

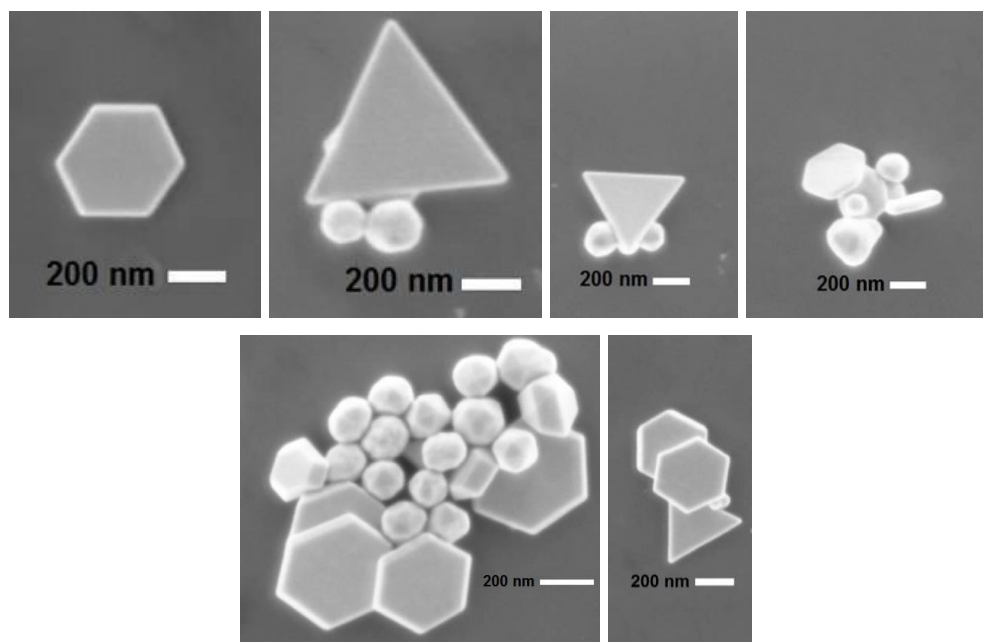

**Figure 5:** Field emission scanning microscope images of gold particles having both geometric anisotropic and distorted shapes synthesized at a processing time of 15 minutes

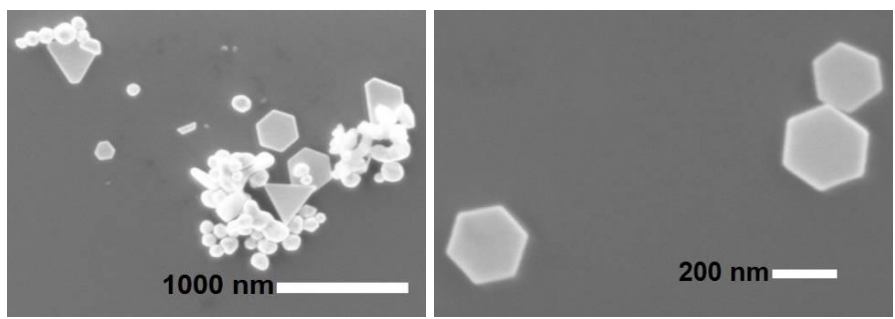

**Figure 6:** Field emission scanning microscope images of gold particles having both geometric anisotropic and distorted shapes synthesized at a processing time of 20 minutes

Here, a tiny particle, in its certain geometry or no specific geometry, is related to the first-hand developed tiny-sized particle (on amalgamation and binding of atoms). Assembling of few such tiny-sized particles resulted into develop the nanoparticles. But the assembling of many such tiny-sized particles resulted into develop the extended shape of their particle. When the assembling of tiny-shaped particles is in a certain order, an extended geometric anisotropic shape of the particle is developed and where this assembling is not in a certain order, a distorted shape of the particle is developed. Further detail on developing different geometric anisotropic particles and distorted particles is given elsewhere [16, 21].

Many tiny-sized particles of a triangular-shape developed when the suitable amount of precursor concentration was utilized [16]. Structure evolution of certain solid nature atoms having different dimension and format was discussed where conservative forces involved to engage the conserved energy for atoms at just above typical-leveled ground surface, at typical-leveled ground surface and at below typical-leveled ground surface [17]. However, in binding of metallic atoms when undertaking the certain transition state at required level of ground, packets (or blocks) of nanoshape energy are placed over their compact monolayer assembly as for the case of present work. Atoms of monolayer assembly bound under the placement of packets of nanoshape energy when they are under re-crystallization state where they undertake diligently the effect of exerting forces working at ground level [18]. A photon energy shape-like a wave is generated while processing the heat energy by a neutral state silicon atom where conservative forces are involved [19]. Atoms eligible to evolve any kind of structure at any scale don't ionize as they are entitled to deal force-energy behaviors since the existence. However, under exceeded propagation of photonic current through the flowing of inert gas atoms having a certain density, they split, and, as a result, the light glow (discharge) known in plasma is appeared [20]. Morphology and structure of tiny-sized particles and large-sized particles were controlled under varying the ratios of pulse OFF to ON time and pulse polarity also [21]; a double-packet nanoshape energy when placed over monolayer assembly, it bound atoms in own shape when they are under the uniform distribution at the solution surface. Efforts were also made to develop triangular-shaped tiny particles in silver and binary composition (silver and gold) under the same setup as in the case of developing gold triangular-shaped tiny particles [22]. Switching morphology-structure of evolving tiny grains, grains and crystallite in carbon films deposited at different substrates has been disclosed [23]. Peaks related to several phases of tiny grains in carbon film indicate a different physical behavior of carbon atoms for each phase [24].

Dissociation of gold atoms is under the supplied energy through the immersed graphitic rod where their uplifting to solution surface is by means of entered forcing energy of travelling photons along with the carried forcing energy (of travelling photons) by the splitted electrons [25]. Because of the fixed precursor concentration, the rate of dissociating gold atoms remained the same for a different processing time of solution. As the amount of precursor remained the same in each experiment along with other parameters, the rate of dissociating gold atoms also remained the same. In the case where processing time was kept 2 minutes, the resulted particles in a large number remained non-faceted. Only few of them were developed faceted under the investigation of a drop of solution out of total quantity of the processed solution ~100 ml. But, in the solution processed for different set durations, the amalgamation of gold atoms to form monolayer assembly at solution surface remained higher at the start of processing solutions. Therefore, the developing of tiny-sized particles in a triangular-shape is not in every case, at initial stage of the process. At shorter process duration, a smaller number of tiny-shaped particles were developed, thus, less triangular-shaped tiny particles were developed, hence, a smaller number of geometric anisotropic shapes of their extended particles developed. The imprecise (inexact) packing of those tiny-sized particles to their developing particles of extended shapes indicates deformation of atoms instead of elongation, which is due to the disordered scheme of atoms at solution surface and at the initial time of processing the solution. This trend gradually becomes favorable in terms of developing triangular-shaped tiny particles for processing the solution for longer period. But, in the case of processing the solution for only 2 minutes, it remained less favorable to develop tiny-sized particles having a triangular-shape. Again, packing of tiny-shaped particles into extended shapes under uniform exertion of forces along with the arriving of those tiny-shaped particles in the loop to pack at centre of light glow remains in less number for 2 minutes processing solution, which results into develop only few particles in geometric anisotropic shapes. In line with that, distorted particles are developed more while processing the solution for shorter time and vice versa for longer time. Because, when processing the solution for longer duration, many of the particles developed in geometric anisotropic shapes as their packed tiny-shaped particles dealt uniform exertion of the forces for assembling each structure of smooth element. Atoms of monolayer assembly at solution surface remained in order where they converted into tiny-sized particles having a triangular-shape and in a higher number. However, the put-forth explanation may be re-visited upto a certain extent when some other amount of precursor concentration is utilized along with different processing unit. Increasing the process time under optimized precursor concentration for fixed amount results into develop the maximum number of triangular-shaped tiny particles [25].

In the packing of tiny-sized particles while developing different extended shapes of particles, adjacent-orientation of electrons in their atoms is not appeared to work uniformly and their tiny-sized particles didn't develop in an equilateral triangular-shape. Instead of perfect packing of such tiny-sized particles, due to geometric constraint, they pack under mixed-behavior of exerting forces, which results into their misfit packing, thus, leaving the afterward packing regions in irregular manner where packing of tiny-sized particles developing a particle deals a further non-uniformity. As shown in the particles developed at the shortest process duration, large tiny-sized particles with an average size bigger than 20 nm deal unfit boundary to maintain packing with respect to the center of developing particle. The tiny-sized particles deal misfit packing with respect to neighboring ones under the uneven adjacent-orientation of electrons in their atoms, however, developing the structure of particles more toward faceted shapes over the time. Such a kind of behavior can be observed in Figure 1 (f) and (g). But the triangular-shaped tiny particles turned a bit slanted due to the sudden stopping of process as they were still in transition to acquire the tempo required to perfectly fit at unfilled regions of developing particle. In the particles where tiny-shaped particles converted atoms of one-dimensional arrays into structures of smooth elements, they packed into the developing shaped-particles under the exertion of uniform forces where they don't indicate unfit packing. Tiny-shaped particles elongated their atoms at uniform rate while exerting the surface force at electron-level in the uniform manner, they order boundaries in uniform manner with respect to the center of a developing particle, thus, envisage lower activation energy. A directly proportional relationship of force-energy was established in atoms of solid behavior when undertaking their typical transition states, which was not the case of gas natured atoms [26]. Based on the nature of electron-dynamics of an atom along with their different modification behaviors, tiny-sized particles targeting certain nanomedicine application may result into the adverse effects not anticipated by the medical specialist [27]. Some recent studies show the applications of nanoparticles in the field of medicine [28, 29] as well as energy [30] when comprised the atoms of certain nature.

In both Figure 1 (f) and Figure 1 (g), tiny-sized particles indicate their misfit packing to develop immature different shape particles. So, it is considerable that these particles kept slightly perturbed packing of tiny-sized particles and might deal the geometric constraint. In any case, the origin of partially adjacent-orientation and partially lateral-orientation of electrons (in atoms of those tiny-sized particles) is under the change of their potential energy, which is slightly under the non-uniform manner. Therefore, electrons of those atoms (of tiny-sized particles) slightly misalign along the poles of exerting forces, which is under the non-orientational based stretching of their clamped energy knots. However, the origin of lateral-orientation of electrons (in atoms of tiny-sized particles) is under the change in their potential energy while exerting the forces of north-south poles (instead of east-west poles) and they undertake expansion and contraction to their clamped energy knots (instead of stretching to clamped energy knots). The origins of adjacent-orientation of electrons in atoms of a tiny-shaped particle is further discussed in a separate study [18] along with their lateral-orientation. Impinging the electron streams of splitting argon gas atoms to the underlying atom (of tiny-sized particle) further alters the behavior of elongation (or deformation) [20].

In SAPR patterns of geometric anisotropic particles, the spotted intensity spots in the patterns are due to forcing energy of reflected photons at the front-surface of structures of smooth elements [31]. Those particles where packings of tiny-shaped particles were more at high-degree angles, the printed intensity spots are more in the form of dots where their center-to-center distance is ~0.24 nm. This center-to-center distance of intensity spots printed in the form of circular dots in SAPR is the same for all particles of multi-dimensional shape. However, those particles where packings of tiny-shaped particles remained more at lower-degree angles, the printed intensity spots are more in the form of lines where their mid-to-mid (instead of center-to-center) distance is ~0.27 nm. This mid-to-mid distance of intensity spots printed in the form of straightened dots in SAPR is expected to be the same for all particles of one-dimensional shape. In SAPR pattern where the base surface belonged to the particle of three-dimensional shape (triangular-shape) or particle of six-dimensional shape (hexagonal-shape), the measured center-to-center distance of orderly spotted dots (~0.24 nm) is less than the one (~0.27 nm) resulted in the case of base surface belonging to the particle of one-dimensional shape (rod- or bar-shaped particles). This difference in the distance of particles of multi-dimensional shape and one-dimensional shape is related to the different width of their structures of smooth elements (and their inter-spacing distance also). Further discussion on the origin of this different distance in one-dimensional shape and multi-dimensional is performed in a separate submission [31].

In the adjacent-orientation, electrons of gold atoms orientate along the east-west poles from their north-south poles (lateral-orientation) where surface forces remain exerting in dominating manner for the opposite-sided tips of each electron. As shown in the Figure 7 (a), an electron of left-side and an electron of right-side from the center of their atom are in the nearly lateral-orientation, which is along the north-south pole.

However, under the exertion (application) of force to downward-sides along opposite poles of those electrons, for the both left-side and right-side to center of atom, they orientate in their nearly adjacent-orientation. As shown in Figure 7 (b), now the electron of left-side to center of atom and electron of right-side to center of the same atom are in their nearly adjacent-orientation, which is along the east-west poles.

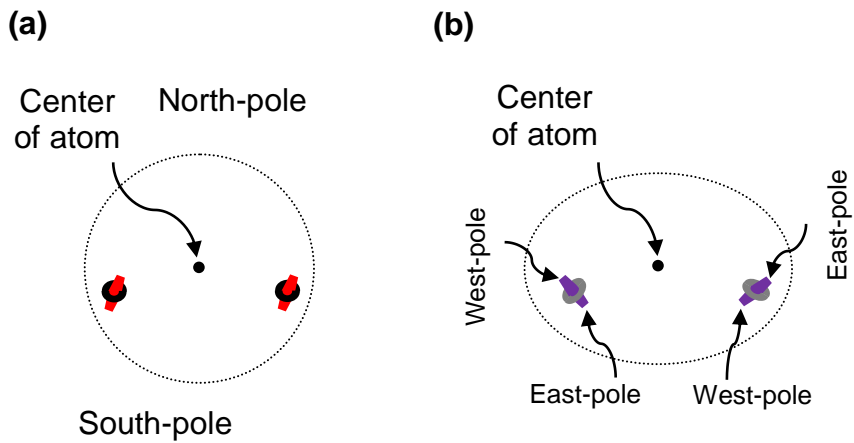

**Figure 7:** Sketch of atom (drawn in estimation) shows only outer electrons of filled states along with center where (a) there is lateral-orientation of left- and right-side electrons and where (b) there is adjacent-orientation of left- and right-side electrons; gold atom indicates electrons of the outer ring only where two included electrons in the center of atom form the zeroth ring following by number of electrons prescribed for electronic configuration, as detailed study of atomic structure belonging to different elements is given elsewhere [26].

A monolayer assembly of gold atoms at solution surface doesn't relate to any specific dimension, orientation or crystallinity. In monolayer assembly, ordering of the atoms is in single layer, they cover region(s) of solution surface by forming the shape more like an arc or a semi-circular shape and at both sides of the center of solution surface (in a standard glass beaker). A layout of the sketch is drawn (in estimation), which is part of another study [25]. At solution surface, atoms are in transition state where their electrons tend to orientate toward the east-west poles and by deviating their orientation of poles from north-south poles. Under the exerted (applied) force, they gradually come into the grip of force prevailing (exerting) at electron-level in the surface format. On placing energy packets of certain shape (under the given protocol) over the uniformly made compact monolayer assembly of gold atoms, they isolate in tiny particles of that shape where the size of those tiny particles is as per availability of energy per unit atom (area) [18].

In line with this, tiny-sized particles of metallic colloids don't develop (evolve) under the forces of van der Waals forces even though van der Waals forces are remained (considered) vital, since a century, in the interactions of atoms and forming aggregates at nanoscale. But, those theories also supported, in many ways, in the accomplishment of many scientific studies at a crucial juncture of scientific research. So, the first author commends the priceless contributions of the Late Dignitaries from the deepest of heart. Respectfully and regretfully, our results don't agree with that theory as all the structural motifs of tiny-sized particles, nanoparticles and particles are subjected to fundamental forces working in different formats of the atomic nature belonging to different elements where engaged (or involved) energy binds atoms and mechanism of binding atoms depends on the nature of their built-in gauge (the scheme of filled and unfilled states of outer ring in an atom and number of rings which is related to distance of the outer ring from the center of the atom). A detail study is given elsewhere [32] discussing the binding mechanisms in different states carbon atoms where both energy and force worked out to evolve the structure; for all sorts of bindings of carbon atoms, they involve energy first, as a result of which, a force is being engaged, which is under the execution of non-confined (and partially confined) inter-state electron-dynamics (instead of under the execution of purely confined inter-state electron-dynamics). In partially confined inter-state electron-dynamics, forces are remained neutral to not influence (exert) a migrating filled state electron to the nearby unfilled one, for example, in structure evolution of one-dimensional graphitic carbon atoms, two-dimensional nanotubes and four-dimensional fullerenes [32]. However, structure evolution in atoms of suitable elements where confined inter-state electron-dynamics are involved is also discussed where force is involved first, because of which, an energy is being engaged in the conserved manner [17]. Therefore, both energy and force contribute to scheme a structure as per the behavior (attained dynamics) of the atom and nature of its built-in gauge of electron-dynamics. So, different tiny-sized particles were synthesized while processing gold solution (or binding of atoms in other suitable class of elements) where bindings of atoms are either due to the involving-engaging of energy-force under the non-conserved manner or due to the involving-engaging of the force-energy under the conserved manner, which describe the very different mechanisms of binding atoms of different elements as compared to those which were explained under the theory of van der Waals forces and associated (linked) theories. Photons of different lengths and energies are discussed elsewhere [19, 20]; a photon is related to packet (entity) of force-energy, which is travelling in the air-medium through forcing its energy from one point to another point. The lattice of different atoms is formed on inter-crossing certain number of photons at the common center where their hollow regions adjust the smallest matter in the form of electrons which are called filled states [26]. The lattice of carbon atom as well as the origins of different states carbon atoms are projected and pinpointed, respectively [32]. These studies do not show any indication of binding atoms through only the force.

The van der Waals forces were originally related to weak forces which enabled the binding of atoms mainly in liquids and solutions [33, 34]. The afterward named (explored) forces, for example, Keesom, Debye and London dispersion forces linked and associated with the weak forces. In our studies, forces are involved also but they involve engaging energy for atoms (amalgamated under appreciable dynamics) to bind when they are in the conservative manner or engage themselves by the involvement of energy for atoms (amalgamated under appreciable dynamics) to bind when they are in the non-conservative manner. They remained neutral where trajectory of involved energy provides the path to transfer the electron from its filled state to nearby unfilled one. Again, those weak forces are belonged to naturally (universally) available forces of poles of earth. Those forces also exist in the liquids and solutions but by introducing the weak effects. Nonlocal van der Waals interactions remained challenging for the theoretical modeling by density functional theory [35]. Therefore, the binding of atoms in any group of elements is by means of energy and force under the reciprocal basis where the physical thing is a matter (at electron-level).

From another perspective, surface plasmons phenomenon is a phenomenon which narrates that, on coupling (trapping) of electromagnetic radiations at air-metal interface, the lattice of nanocrystal or large crystal starts oscillating collectively. However, as observed in Figure 4 (b) and (c), the structures are affected in different ways and where atoms suitably elongated, their structures converted into structures of smooth elements having the inter-spacing distance ~0.12 nm as shown in Figure 4 (d). However, in Figure 4 (c), the structure of smooth element measures width ~0.21 nm, which was resulted under the attachment of two structures of smooth elements where electrons and unfilled energy knots of inner sides of elongated atoms (of structures of smooth elements having width ~0.12 nm in each) coordinated to bind under bearing the opposite force at their normal. This width is almost twice to that recorded in the case of elongated atoms of one-dimensional arrays in particles of multi-dimensional shapes where their structures of smooth elements do not coincide (attach) forcefully and maintain the structure under the availability of exerting forces at normal (natural) grounds. These results agree to the observations that there is no such phenomenon of localized surface plasmon polaritons or surface plasmons. Again, the interaction of photons while traveling along the interface (surface) of tiny-sized particles doesn't enable any sort of collective oscillation of the lattice as it falls under the largely studied phenomenon –a surface plasmons phenomenon. In fact, the traveling photons along the interface further shape the elongated atoms of one-dimensional arrays of tiny-sized particles. Structures of smooth elements are related to elongated atoms of one-dimensional arrays forming the monolayer tiny-sized particle. In each layer of elongated atoms, electrons undertake adjacent-orientation where they are aligned side-to-side from the center of their atom but those electrons which are at the ends assembled (bound) by the side-to-side atoms under orientational-based stretching of clamped energy knots, thus, developing a structure of smooth element [18]. So, the process of converting atoms of one-dimensional array into structure of smooth element is a completely different behavior to the ones which are described in terms of dipole-dipole interactions as they infer attraction and repulsion within opposite charge poles and alike charge poles, respectively. Further details of the developing structures of smooth elements in triangular-shaped tiny particle (of monolayer) along with the further shaping of elongated atoms of one-dimensional arrays through forcing energy of travelling photons are discussed elsewhere [18]. Again, single atom embedded by monolayer lattice doesn't deal (undertake) any sort of collective oscillation where it further aligned the elongation behavior under the impact of forcing energy of travelling photons aside to it as discussed elsewhere [20].

Tiny-sized particles didn't shape in a triangular-shape where their atoms are diffusing under the elongation and deformation behaviors having orientation of the electrons along both uni-direction and multi-direction to the maximum extent. For uni-direction, elongation behavior of the atoms (of tiny-sized particle) is taken place where their electrons disrupt to undertake the adjacent-orientation and under the orientational-based stretching of their clamped energy knots. This is because of the exertion of force to the opposite poles (of electrons) along the single axis but along the both sides from the centers of their atoms. For multi-direction, deformation behavior of some of the atoms (of tiny-sized particle) is taken place where their electrons disrupt to undertake the mixed-orientation and under the non-orientational-based stretching (and compression) of their clamped energy knots. This is because of the exertion of force to electrons in non-uniform manner as they gained potential energy at different level. Electrons of both left-side and right-side to center of their atom do not orientate along the single axis (uni-direction) to elongate it as they possess the variable potential energy. So, when the atom is not elongating, this is because of the exertion of force to the opposite poles (of electrons) along the multiple axes but along the both sides from the centers of their atoms. In Figure 1 (f) and Figure 1 (g), this diffusion-orientated phenomenon of electrons in (elongated and deformed) atoms of non-regular (geometrically constrained) tiny-sized particles is obvious where their (loose) packings develop (imprecise/inexact) mono-layers of immaturely developed particles.

As shown in Figure 7 aligning of electrons along uni-direction for their atoms (either for lateral-orientation or for adjacent-orientation) by remaining within occupied states provide channelized inter-state gaps where given field of photonic current exceeded the acceleration of travelling (or propagating) through channel of smooth and uniform inter-state electron gaps. So, anisotropic nanoparticles (or particles) describe high photonic properties. However, the anisotropic nanoparticles (or particles) of multi-dimensional shapes (hexagonal-, pentagonal- and triangular-shapes) indicate different potential for applications than the anisotropic nanoparticles (or particles) having one-dimensional shapes (rod- or bar-shapes). Because, the nanoparticles (or particles) of multi-dimensional shapes and one-dimensional shapes possess the different inter-state electron gaps [31]. Photon-induced nanoparticles (or particles) other than anisotropic behaviors possess potential for the catalytic properties as their atoms don't entertain inter-state electron gaps uni-directionally in the long length, area or volume. The field is remained in the blockage from different regions of photon-induced nanoparticles (or particles) and that field (force-energy) is being utilized locally to kill bacteria, to clean water or for other useful purposes. For the case of distorted-/spherical-shaped nanoparticles (or particles), they can be the strong candidate under replication of light (field) through sun (input source) depending on the number of their contained atoms having the active mode of electron-dynamics. However, the potential use of highly-anisotropic nanoparticles (or particles) is transforming the optimized usage of field from their one (input) end (point of supply field) to other (output) end (point of utilizing field).

A crystalline structure infers that atoms of a structure belonging to few nanometers area have the ordering of a same orientation. In the case where atoms of a structure cover the longer-length area in their ordering of a same orientation, it is also a crystalline structure but of a large area. An anisotropic nanoparticle/particle infers its dimension where each dimension should adhere the same structure (ordering of the atoms); a triangular-shaped particle has three-dimensions, hexagonal-shaped six-dimensions and pentagonal-shaped five-dimensions where the structure of each face (along each dimension) is a crystalline structure as the ordering of the atoms is in the same fashion where atoms obey the same orientation.

A multi-crystalline nanoparticle (or particle) is not related to anisotropic nanoparticle or particle but related to distorted nanoparticle (or particle) where some atoms are in different orientation as compared to orderings of atoms having different orientations in the neighboring regions. However, an isotropic nanoparticle (or particle) is related to purely sphere-shape where circular distribution/ordering of its atom occurs. A pentagonal-shaped nanoparticle shown in Figure 4 is related to multi-dimensional shape having five orientations of its faces where each face carries the same crystalline structure but for the different zones of the nanoparticle. However, assembling structures of smooth elements for each face of that pentagonal-shaped particle (nanoparticle) is under a bit slanted orientation, which is related to arisen stress called twin boundaries and due to slightly misfit packing of their tiny-shaped particles. In fact, arriving of tiny-shaped particles from five different regions of solution surface to assemble their structures of smooth elements at common center develops a constraint because the applied (fundamental) forces exerting at electron-level loses their symmetry at the assembling point having common center. But, they are well-functioning (exerting) at electron-level as the disturbance-level of packing of tiny-shaped particles and assembling their structures of smooth elements in developing triangular- and hexagonal-shaped nanoparticles (or particles) are minimized.

In view of above-said, the origin of a diffusion mechanism for either deformed atom or elongated atom of a tiny-sized particle is based on the dedicated-orientation of their relevant electrons and to the maximum extent for either multi-direction or uni-direction. Electrons of the atoms introducing the diffusion mechanism is as per their pulling or pushing (controlled-orientation with respect to clamped energy knots) by remaining inside the clamped energy knots. Pulling of the electron deals (undertakes) the stretching of clamped energy knot but pushing of the electron deals (undertakes) the compression of clamped energy knot. The rate of diffusing atoms (so, their tiny-sized particles or grains) should be based on the collective pulling or pushing of electrons in them either in uni-direction or in multi-direction. As per supplied heat energy, electrons of atoms are being orientationally-controlled under the gained potential energy either for the fixed orientation at discrete potential energy or for the different orientations at variable potential energy. At this node, the concepts of magnetism can be re-investigated in atoms of suitable classes along with description of the Coulomb's law (and concepts of electricity) in new insights. But, in diffusion mechanism, electrons deal infinitesimal displacements (while remaining clamped by their energy knots) at very slow pace under the absorbed heat energy by their atoms as the minute element of the force is around them. This is not the case when considering the magnetization process in suitable materials at bulk level and Coulomb's law in suitable materials at atomic (nano) level as the direct element of the force is involved at very high pace without involving the heat.

When atoms of monolayer assembly at solution surface assemble (bind) into triangular-shaped tiny particles on placing a packet of nanoshape energy, atoms of their one-dimensional arrays elongate one-dimensionally under the stretching of energy knots clamped electrons. The tiny-shaped particles pack to develop both one- and multi-dimensional extended shapes where their structures of smooth elements assemble at the centre of light glow by arriving from different regions of the solution surface [25]. For the case of one-dimensional shape of the particles, atoms of triangular-shaped tiny particles elongated less, thus, developing more width in each structure of smooth element [31]; in the case of packing of tiny-shaped particles (to develop them) in shape-like triangular or hexagonal, a similar mechanism is involved, as for the case of developing rod- or bar-shaped particles but their arrival to pack at centre of light glow is from the different zones of the solution surface.

**Conclusions:**

The present work describes the development mechanisms of different tiny-sized particles and their extended shapes under a different processing time in pulse-based electron-photon-solution interface process. Under lengthy process time of solution where precursor concentration was 0.60 mM, a greater number of tiny-sized particles and their extended shapes develop having the geometric shapes. An anisotropic particle of multi-dimension having high aspect ratio develops nearly in sub-millisecond/millisecond time. Non-uniform adjacent-orientations of electrons to atoms of tiny-sized particles result into misfit packing for their large-sized particles. The loosely packed tiny-sized particles acquire smoothness for developing their extended shapes over the time as per diffusing rate of electrons belonging to their uni-directional ordered atoms where the stretching of lattices (energy knots net) is orientational-based (along the same axis). The particles developed at 15 minutes and 20 minutes process durations are highly-faceted showing the smooth features of their surfaces where no sign of misfitly packed tiny-sized particles is observed. Scanning microscope images of particles show identical features as for the case of transmission microscope. At fixed suitable precursor concentration, increasing the processing time of solution upto certain duration increases the number of tiny particles in a triangular-shape, hence, their extended shapes also.

For particles developed at a different processing time, the inter-spacing distance of structures of smooth elements and width of a structure of smooth element are remained the equal (same), but they are different in the case of particles of one-dimension and multi-dimension. The inter-spacing distance along with width of structure of smooth element becomes ~0.21 nm when two parallel structures of smooth elements (each having width ~0.12 nm) coincide (adhere) under the exerting force oppositely to their normal where certain electrons of elongated atoms (from their inner sides) bind by the certain unfilled energy knots of elongated atoms (from their inner sides) belonging to their structures of smooth elements.

In visualizing the structure of particles, the spotted dots are related to the reflected photons at the surface under observation, which are not due to the diffraction of electrons. Adjacent-orientation of electrons in atoms of tiny-sized particles (or large-sized particles/extended shapes) is under the orientational-based stretching of clamped energy knots as per exerting surface force along their opposite poles and from the centers of their atoms. Atoms of tiny-sized particles (grains) diffuse (migrate) at the rate of orientation of electrons along the single axis (uni-direction) from their naturally occupied orientations in clamped energy knots. Hence, origin of diffusing the matter at electron-level for their atoms (and tiny-sized particles) solicits re-visiting of diffusion-based laws and theories.

Our investigations don't comply with binding mechanisms as covered largely under van der Waals interactions. Different sorts of forces are engaged under the involvement of packets of nanoshape energy where the binding of gold atoms in monolayer triangular-shaped tiny particle occurs if the distribution of atoms in the monolayer assembly is compact. Binding of gold atoms under the packets of certain nanoshape energy only engages forces to attain appropriate coincidences where they maintained the orientation of their electrons as per exertion of forces, so, in the case of their tiny-sized particles and extended shapes also.

Our investigations don't agree with that i.e., a tiny-sized lattice of atom starts its collective oscillation on trapping the light. On the traveling photons having suitable wavelength aside to transitional state atoms of one-dimensional arrays belonging to monolayer tiny-sized particle, they further shape them under transported forcing energy, thus, converting (modifying) them into improved (flat) featured structures of smooth of elements. This clearly negates the largely studied phenomenon known in surface plasmons phenomenon. Our study presents the clear clues to address electricity and magnetisms under the new insights.

The systematic cycle of developing different extended shapes starting from gold monolayer assembly at solution surface following by developing tiny-sized particle is the hallmark to explore several scientific breakthroughs and under the control of process time for their different conditions. These investigations enlighten us that atoms belonging to different elements require their re-investigation not only at their own scale, but at nanoscale and bulk scale also and then interactions to different profitable counterparts.

**Acknowledgements**

We thank National Science Council now Ministry of Science and Technology, Taiwan (R.O.C.) for awarding postdoctorship: NSC-102-2811-M-032-008 (August 2013- July 2014) and appreciate the support of Dr. Kamatchi Jothiramalingam Sankaran, National Tsing Hua University, Taiwan (R.O.C.) for helping in materials' microscopy. Dr. Ali also acknowledges support of Dr. M. Ashraf Atta while writing article.

## Authors' biography:

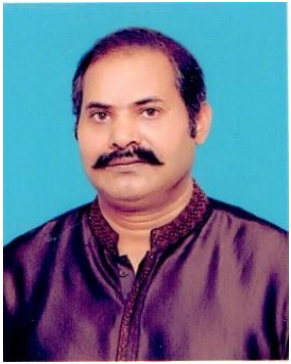

Mubarak Ali graduated from University of the Punjab with B.Sc. (Phys& Maths) in 1996 and M.Sc. Materials Science with distinction at Bahauddin Zakariya University, Multan, Pakistan (1998); thesis work completed at Quaid-i-Azam University Islamabad. He gained Ph.D. in Mechanical Engineering from Universiti Teknologi Malaysia under the award of Malaysian Technical Cooperation Programme (MTCP;2004-07) and postdoc in advanced surface technologies at Istanbul Technical University under the foreign fellowship of The Scientific and Technological Research Council of Turkey (TÜBİTAK; 2010). He completed another postdoc in the field of nanotechnology at Tamkang University Taipei (2013-2014) sponsored by National Science Council now M/o Science and Technology, Taiwan (R.O.C.). Presently, he is working as Assistant Professor on tenure track at COMSATS University Islamabad (previously known as COMSATS Institute of Information Technology), Islamabad, Pakistan (since May 2008) and prior to that worked as assistant director/deputy director at M/o Science & Technology (Pakistan Council of Renewable Energy Technologies, Islamabad; 2000-2008). He was invited by Institute for Materials Research, Tohoku University, Japan to deliver scientific talk. He gave several scientific talks in various countries. His core area of research includes materials science, physics & nanotechnology. He was also offered the merit scholarship for the PhD study by the Government of Pakistan, but he couldn't avail. He is author of several articles available at links; https://scholar.google.com.pk/citations?hl=en&user=UYjvhDwAAAAJ, https://www.researchgate.net/profile/Mubarak_Ali5.

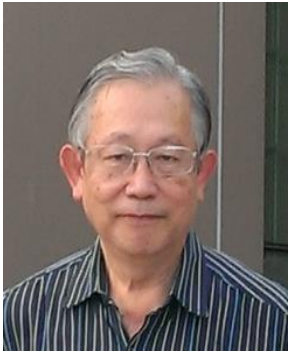

I-Nan Lin is a senior professor at Tamkang University, Taiwan. He received the Bachelor degree in physics from National Taiwan Normal University, Taiwan, M.S. from National Tsing-Hua University, Taiwan, and the Ph.D. degree in Materials Science from U. C. Berkeley in 1979, U.S.A. He worked as senior researcher in Materials Science Centre in Tsing-Hua University for several years and now is faculty in Department of Physics, Tamkang University. Professor Lin has more than 200 referred journal publications and holds top position in his university in terms of research productivity. Professor I-Nan Lin supervised several PhD and Postdoc candidates around the world. He is involved in research on the development of high conductivity diamond films and on the transmission microscopy of materials.